\documentclass[12pt]{emulateapj}

\newcommand{\jybeam}{Jy beam$^{-1}$}

\newcommand{\opacity}{cm$^2$ g$^{-1}$}

\shorttitle{Properties of the TWA 7 Disk}
\shortauthors{Matthews et al.}
\begin{document}

\title{Mass and Temperature of the TWA 7 Debris Disk}

\author{Brenda C.~Matthews}
\affil{Herzberg Institute of Astrophysics, National Research Council
  of Canada, 5071 West Saanich Road, Victoria, BC, V9E 2E7, Canada}
\email{brenda.matthews@nrc-cnrc.gc.ca}

\author{Paul G.~Kalas}
\affil{Department of Astronomy, University of California,
  Berkeley, CA, 94720-3411, U.S.A.}
\email{kalas@astron.berkeley.edu}

\and

\author{Mark C.~Wyatt}
\affil{Institute of Astronomy, University of Cambridge, Madingley
  Road, Cambridge, CB3 0HA, U.K.}
\email{wyatt@ast.cam.ac.uk}

\begin{abstract}
We present photometric detections of dust emission at 850 and 450
\micron\ around the pre-main sequence M1 dwarf TWA 7 using the SCUBA
camera on the James Clerk Maxwell Telescope.  These data confirm the
presence of a cold dust disk around TWA 7, a member of the TW Hydrae
Association.  Based on the 850 \micron\ flux, we estimate the mass of
the disk to be 18 $M_{\rm lunar}$ (0.2 $M_\oplus$) assuming a mass
opacity of 1.7 \opacity\ with a temperature of 45 K.  This makes the
TWA 7 disk ($d=55$ pc) an order of magnitude more massive than the
disk reported around AU Microscopii (GL 803), the closest (9.9 pc)
debris disk detected around an M dwarf.  This is consistent with TWA 7
being slightly younger than AU Mic. We find that the mid-IR
and submillimeter data require the disk to be comprised of dust at a
range of temperatures.  A model in which the dust is at a single
radius from the star, with a range of temperatures according to grain
size, is as effective at fitting the emission spectrum as a model in
which the dust is of uniform size, but has a range of temperatures
according to distance. We discuss this disk in the context of known
disks in the TW Hydrae Association and around low-mass stars; a
comparison of masses of disks in the TWA reveals no trend in mass or
evolutionary state (gas-rich vs.\ debris) as a function of spectral
type.

\end{abstract}


\section{Introduction}

TWA 7 (2MASS J10423011-3340162, TWA 7A) is a weak-line T Tauri star
identified as part of the TW Hydrae Association \citep[TWA,][]{kas97}
by \cite{web99} based on proper motion studies in conjunction
with youth indicators such as high lithium abundance, X-ray activity
and evidence of strong chromospheric activity.  Disk systems were
inferred around four of the 18 association members \citep{zuc04} from
measurements of IR excess with the {\it Infrared Astronomical
Satellite (IRAS)}.  TW Hydra itself (a K7 pre-main sequence star)
hosts the nearest protostellar disk to the Sun.  Another accreting
disk is observed around one member of the triple system Hen 3-600
\citep{muz00}, and two debris disk systems have been detected, around
the A0 star HR 4796A \citep{jur91,sch99} and one of two spectroscopic
binary components of the quadruple system HD 98800
\citep{jay99,geh99}, which is a K5 dwarf.  TWA 7 was not detected by
{\it IRAS}.

Based on the width of the Li 6707 \.{A} line, \cite{neu00} deduced
that TWA 7 is a pre-main sequence star.  TWA 7 was not detected by
Hipparcos, but its membership in the TWA sets its distance to be $55
\pm 16$ pc \citep{neu00,wei04,low05}.  Its spectral type is M1 based
on LRIS spectra \citep{web99}.  Based on existing photometry,
evolutionary tracks and isochrone fitting, \cite{neu00} derived an age
of 1-6 Myr (i.e., roughly coeval with other TWA stars) and a mass of
0.55 $\pm$ 0.15 $M_\odot$.  The age of the association is generally
taken to be $\sim 8-10$ Myr \citep{sta95,zuc04}.  This is the age when
planet formation is thought to be ongoing and when disk dissipation is
occurring. Thus, the TWA is the ideal cluster in which
to observe the transitions from pre-main sequence stars with
proto-planetary (gas-rich) disks to main sequence stars with debris
(gas-poor) disks.

TWA 7 has been observed for infrared excess emission several
times. The presence of a disk around TWA 7 was first noted in
submillimeter observations by \cite{web00}.  They measured a flux of
$15.5 \pm 2.4$ mJy at 850 \micron. Neither \cite{jay99} nor
\cite{wei04} detected any excess associated with TWA 7 in the
mid-infrared.  However, \cite{low05} report detections of IR excess at
24 and 70 \micron\ toward TWA 7 with the Spitzer Space Telescope.
Based on these data and existing shorter wavelength data on the
stellar photosphere, \cite{low05} derive a disk temperature of 80 K
and a lower limit to the mass of $2.4 \times 10^{23}$ g ($3.3 \times
10^{-3}$ $M_{\rm{lunar}}$) for the disk.  This is a lower limit
because the 70 \micron\ data are not sensitive to colder material in
the outer disk and dust grains exceeding a few hundred micron in size.
A search for substellar companions using the NICMOS chronograph on
HST did not reveal any evidence of the disk \citep{lowr05}; a nearby
point source is identified as a background object.

The study of debris disks around members of an association permits the
study of the evolution of disks as a function of spectral type alone,
since the disks likely formed coevally and with similar compositions.
It is also possible to judge whether the presence and evolution of
disks around multiple stellar systems is comparable to those around
single stars. The discovery of disks around low-mass stars is
relatively recent \citep{gre98,liu04,kal04}.  The low radiation field
of late-K and M dwarfs means that the disks are faint in scattered
light compared to disks around more massive stars, and hence they were
less frequently targeted by scattered light searches for disks.
However, scattered light imaging is often a follow-up technique used
after an infrared excess has been discovered \citep[e.g., AU
Mic,][]{kal04}.  In fact, the low radiation fields may favor
long-lived and slowly evolving disks, since some disk material may be
unchanged from the original proto-planetary disks
\citep[e.g.,][]{gra07}.  Spitzer has detected evidence of disks around
a few K stars \citep{che05,bry06,uzp05,gor04,bei05}, but \cite{bei06}
note that in a sample of 61 K1 to M6 stars, no excess emission is
detected at 70 \micron.  This is well below the expected detection
rate if the disk fraction is at all comparable to the $\sim 15$\%
observed around solar-type stars. The discovery of disks around K- and
M-type dwarfs may be difficult at far-infrared wavelengths because
material at similar radii to disks around early-type stars will be
cool and will not radiate sufficiently. However, submillimeter
observing sensitivities and the dust temperature conspire to allow
these objects to be discovered at submillimeter wavelengths (Zuckerman
2001). Submillimeter observations are sensitive to colder, larger
grains, which are more likely to be optically thin than the warmer
far-infrared emitting dust \citep{hil88}.

We report here detections of submillimeter excess emission at 450 and
850 \micron\ around TWA 7 using the James Clerk Maxwell Telescope
(JCMT). In $\S$ \ref{obs}, we summarize our observations. In $\S$
\ref{res}, we present our results.  We discuss the relevance of these
data to the TWA and the population of disks around low-mass stars in
$\S$ \ref{disc}; our results are summarized in $\S$ \ref{conc}.

\begin{deluxetable*}{lcrl}[hb!]
\tablecolumns{4} 
\tablewidth{0pc} 
\tablecaption{Fluxes of TWA 7}
\tablehead{\colhead{Wavelength} & \colhead{Magnitude} & 
\colhead{Flux} & \colhead{Reference}  \\
\colhead{[$\mu$m]} & & \colhead{[mJy]} & }
\startdata 
0.44 (B) & 12.55 & 39.4  & HST Guide Star Catalog \citep{las96} \\
0.44 (B) & 12.3 & 49.7 &  USNO-A2.0 \citep{mon98} \\
0.54 (V) & 11.06 & 142.4 & reported in \cite{low05} \\
0.64 (R) & 11.2 & 97.4 & USNO-A2.0 \citep{mon98} \\
1.25 (J) & 7.78 & 1259.5 & \cite{web99} \\
1.65 (H) & 7.13 & 1476.3 & \cite{web99} \\
2.16 (Ks) & 6.90 & 1159.1 & 2MASS PSC \\
2.18 (K) & 6.89 & 1148.8 & \cite{web99} \\
12 & -- & 70.4 $\pm$ 8.6 & \cite{wei04} \\
24 & -- & 30.2 $\pm$ 3.0 & \cite{low05} \\
70 & -- & 85 $\pm$ 17 & \cite{low05} \\
450 & -- & 23 $\pm$ 7.2 & this work \\
850 & -- & 9.7 $\pm$ 1.6 & this work \\
\enddata
\tablecomments{Conversion from magnitudes to Janskys has been done using
  zero points from {\it Allen's Astrophysical Quantities}
  \citep{cox99}. The 12 \micron\ estimate by \cite{wei04} agrees with
  that of \cite{jay99} to within 1 $\sigma$ and is consistent with the
photospheric flux expected from TWA 7.}
\label{fluxes}
\end{deluxetable*}

\section{Observations and Data Reduction}
\label{obs}

Observations were made in 2004 October 19 using the photometry mode on
the Submillimeter Common User Bolometer Array (SCUBA) on the JCMT
\citep{hol99}.  The pointing center of the observation was $\alpha =
10^{\rm h}42^{\rm m}$30\fs3, $\delta = -33$\degr40\arcmin16\farcs9
(J2000).  The on-source integration time was 1.6 hours.  Flux
calibration was done using Mars, yielding flux conversion factors
(FCFs) of 289.2 $\pm$ 1.4 \jybeam\ volt$^{-1}$ at 850 \micron\ and
367.5 +/- 15 \jybeam\ volt$^{-1}$ at 450 \micron.  The absolute flux
calibration is accurate to $\sim 20-30$\%. Pointing was checked on the
source $1034-293$.  The weather was excellent during the observations,
with a CSO tau measurment at 225 GHz of $\sim$ 0.04.  The extinction
was corrected using skydips to measure the tau at 850 and 450 \micron.
The mean tau value was 0.15 at 850 \micron\ in four skydips and 0.59
at 450 \micron\ in three skydips.  The extinction values derived from
the skydips were consistent with the values extrapolated from the CSO
tau values according to the relations of \cite{arc02}.

The data were reduced using the Starlink SURF package \citep{jen98}.
After flatfielding and extinction correction, we flagged noisy
bolometers rigorously; eleven (of 37) bolometers were removed in the
long wavelength array and 26 (of 91) were removed from the short
wavelength array.  The photometry data were then clipped at the 5
$\sigma$ level to remove extreme values. Short timescale variations in
the sky background were then removed using the mean of all bolometers
except the central one in each array.  The average and variance were
then taken of each individual integration for the central bolometers
of the long and short wavelength arrays after a clip of 3 $\sigma$ was
applied.  In the case of the 450 \micron\ data, the signal-to-noise
ratio was improved by applying a subsequent $2 \sigma$ clip to the
remaining data.

\section{Results}
\label{res}

\begin{figure}[h!]
\vspace*{5cm}
\includegraphics{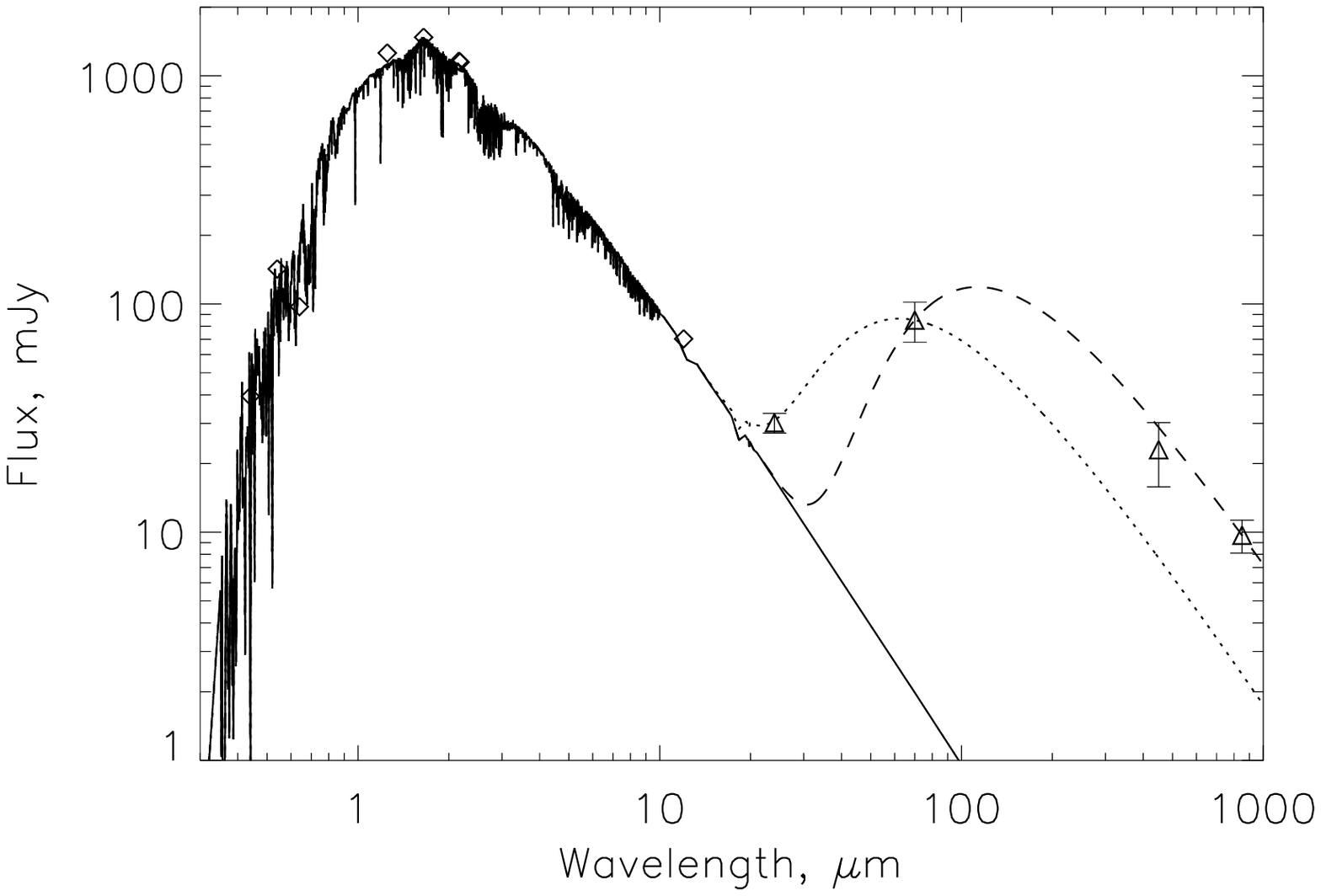}
\caption{The spectral energy distribution of TWA
  7.  The optical and near-infrared data ({\it diamonds}) are modeled
  with a NextGen model \citep{hau99} scaled to the mass and luminosity
  of TWA 7 \citep{low05}.  The star has a temperature of $\sim 3500$ K
  \citep[best fit to stellar photometry,][] {low05,web99}. Triangles
  mark detections from the Spitzer Space Telescope and submillimeter
  detections from the JCMT.  These fluxes show clear excess when
  compared to the stellar photosphere.  The flux values are reported
  in Table \ref{fluxes}.  Fits for two single temperature blackbodies
  (parameters are described in Table \ref{modelparam}) to the TWA 7
  data show that no single temperature fits all four measurements of
  excess emission. An 80 K blackbody, Model 1, ({\it dotted line}) fits the 24
  and 70 \micron\ data \citep{low05}, but underestimates the
  submillimeter fluxes.  The 70, 450 and 850 \micron\ data are well
  fit by a 45 K blackbody, Model 2, ({\it dashed line}), but a disk this cold
  cannot account for the observed 24 \micron\ excess.}
\label{sed_1temp}
\end{figure}

We have detected emission toward TWA 7 at both 850 \micron\ and 450
\micron.  The fluxes measured are 9.7 $\pm$ 1.6 mJy ($6.1 \sigma$) and
23.0 $\pm$ 7.2 mJy ($3.2 \sigma$), respectively.  Errors are
statistical, and do not include the typical flux uncertainty of $\sim
20-30$\% for submillimeter single-dish calibration.  Utilizing the
measured fluxes of the star (Table \ref{fluxes}) and recently
published Spitzer data \citep{low05}, we construct the spectral energy
distribution (SED) for this source (Figure \ref{sed_1temp}).  The
submillimeter data clearly represent an excess of emission over the
photospheric emission from TWA 7, as detected at 70 \micron\ by
\cite{low05}.

The flux we measure at 850 \micron\ is only about 70\% that measured
by \cite{web00}.  Taking into account the typical $20-30$\%
uncertainty in absolute flux calibration between epochs, the fluxes
become $15 \pm 3.8$ mJy (2000) and $9.7 \pm 2.5$ mJy (2004), which are
consistent within the errors.  The instability of flux calibration in
the submillimeter is well known, and may be assuaged somewhat by the
ability to flux calibrate more often with SCUBA-2, the next generation
submillimeter camera on the JCMT.  In the interests of consistency
with the only 450 \micron\ detection in this work, we adopt the 850
\micron\ flux from the later epoch and do not attempt to combine the
two datasets.

\subsection{Submillimeter Excess and Temperature}
\label{3.1}

\cite{low05} derive a disk temperature of 80 K for TWA 7 based on its
infrared excess values at 24 and 70 \micron.  The temperature of the
star is derived to be 3500 K, which is consistent with the
log($T_{{\rm eff}}$) of 3.56 reported by \cite{neu00} based on the
stellar SED.

Figure \ref{sed_1temp} shows the flux density distribution toward TWA
7.  The stellar fluxes and fit to the stellar photosphere are taken
from \cite{low05} based on NextGen models by \cite{hau99} using a grid
of \cite{kur79}.  The submillimeter excesses are evident when compared
with the stellar spectra.  Based on their detection of TWA 7 in the
mid-infrared, \cite{low05} determine that the dust orbiting TWA 7
exists at radii $\ge 7$ AU from the star and has a temperature of 80
K.  We have constructed several models for dust emission around a 0.55
$M_\odot$ star (see Table \ref{modelparam} for their parameters). We
present two single temperature models in Figure \ref{sed_1temp},
neither of which is capable of fitting the measured excess fluxes at all four
wavelengths. Model 1 (dotted line in Fig.\ \ref{sed_1temp}) is that of
\cite{low05}: a blackbody of 80 K, which fits the 24 and 70 \micron\
data, but cannot reproduce the fluxes at longer wavelengths.  Model 2
is a colder black body at a temperature of 45 K (dashed line of Fig.\
\ref{sed_1temp}), which fits the three longest wavelength fluxes, but
cannot reproduce the 24 \micron\ flux. Thus we conclude that the dust
in this disk cannot be at a single temperature; rather there are a
range of temperatures responsible for the observed emission.

\begin{deluxetable*}{cccccccc}[b!]
\tablecolumns{7} 
\tablewidth{0pc} 
\tablecaption{Dust Model Parameters}
\tablehead{\colhead{Model} & \colhead{Plotted as} & \colhead{Temperature} & 
  \colhead{Dmin} & \colhead{Dmax} & \colhead{q} & \colhead{Radius} &
  \colhead{Mass ($M_\oplus$)}}
\startdata 
1 & dotted Fig.\ \ref{sed_1temp} & 80 K & -- & -- & -- & -- & 0.025 \\ 
2 & dashed Fig.\ \ref{sed_1temp} & 45 K & -- & -- & -- & -- & 0.2 \\
3 & dotted Fig.\ \ref{sed} & 21 - 65 K & 0.9 \micron\ & 1 m & 1.78 & 100 AU & 6.0 \\
4 & dashed Fig.\ \ref{sed} & $> 38$ K & -- & -- & -- & $\le 35$ AU & 0.2 \\
\enddata
\label{modelparam}
\end{deluxetable*}

\begin{figure}[h!]
\vspace*{5cm}
\includegraphics{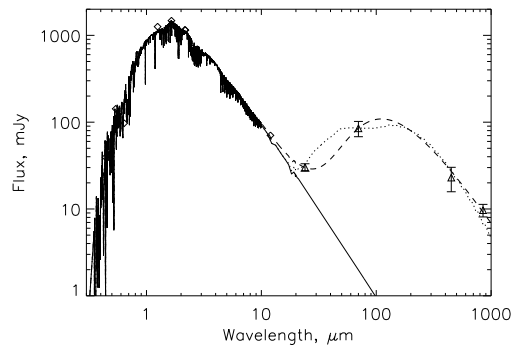}
\caption{The measured SED for TWA 7. Symbols and
  fit to the stellar spectrum is as for Fig.\ \ref{sed_1temp}. Two
  multi-temperature models are fit to the excess emission arising from
  the disk. These are described in detail in the text, and the parameters are
  summarized in Table \ref{modelparam}.  In Model 3 ({\it dotted line}),
  different temperatures arise due to grains of different sizes
  located at a common radius from the star; while Model 4 ({\it dashed
  line}) contains grains distributed at a range of distances from the
  star. }
\label{sed}
\end{figure}

Two models that are able to fit the observed emission spectrum are
shown in Figure \ref{sed}. In both models the dust has a range of
temperatures, however there are two different physical motivations
behind the origin of this range.  

In Model 3 (dotted line of Fig.\ \ref{sed}), the dust is assumed to
all lie at the same distance from the star, $r_0$. It is assumed to
have a range of sizes with a power law distribution defined by the
relation $n(D) \propto D^{(2-3q)}$ for grains of size $D$.  This power
law is assumed to be truncated below dust of size $D_{\rm min} = 0.9$
\micron, the size for which $\beta = F_{\rm rad}/F_{\rm gra} = 0.5$
for compact grains.  The dust is assumed to be composed of compact
spherical grains of a mixture of organic refractories and silicates
\citep{li97}, and interaction with stellar radiation determines the
temperature that dust of different sizes attains. The same model was
used in \cite{wya02} to model the emission from the Fomalhaut disk,
where more detail can be found on the modelling method.  The emission
spectrum could be fitted by dust at a radius $r_0 = 100$ AU with a
size distribution described by $q=1.78$. The temperatures of dust in
this model range from 21 K for the largest grains ($>100$ \micron) to
65 K for the smallest grains (0.9 \micron).

\begin{figure}[h!]
\vspace*{5cm}
\includegraphics{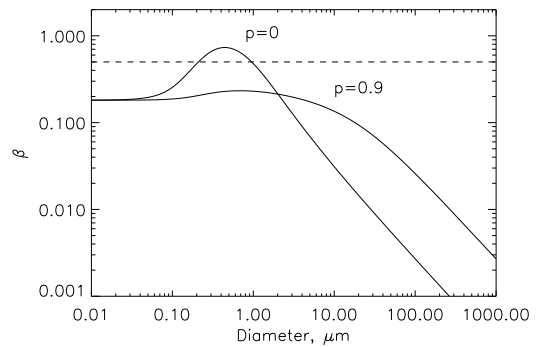}
\caption{Ratio of the radiation force acting on dust
grains of different size in the TWA 7 disk to the force of stellar
gravity, $\beta=F_{rad}/F_{gra}$. Two different models are shown:
compact grains ($p=0$) and porous grains ($p=0.9$). Dust with
$\beta>0.5$ is unbound from the star as soon as it is released from a
planetesimal. This figure demonstrates that radiation pressure is not
a significant mechanism to remove dust grains from the TWA 7 system
and is particularly ineffective if the grains are highly porous.}
\label{betas}
\end{figure}

The size distribution used in Model 3 is close to that expected in a
collisional cascade wherein dust is replenished by collisions between
larger grains, since this results in a size distribution with $q \sim
1.83$ and would be truncated below the size of dust for which
radiation pressure would place the dust on hyperbolic orbits as soon
as they are created. The effect of radiation forces on small grains is
quantified by the parameter $\beta =F_{\rm rad}/F_{\rm gra}$ (which is
not to be confused with the index of dust emissivity of grains that
moderates the Rayleigh-Jeans tail of the SED of cold dust), and it is
dust with $\beta>0.5$ which is unbound. However, due to the low
luminosity of M stars, it is not clear whether radiation pressure is
sufficient to remove dust grains from the disk system. Figure
\ref{betas} shows $\beta$ as a function of dust grain size for dust
around TWA 7 ($M_* = 0.55 \ M_\odot$, $L_* = 0.31 \ L_\odot$). While
$\beta$ is larger than 0.5 for compact ($< 0.9$ \micron) grains, this
condition is only met for a narrow region of the size distribution.
Furthermore, if the dust grains are porous, as around AU Mic
\citep{gra07}, then no grains will have
$\beta > 0.5$.  An alternative origin for the
small grain cut-off could be stellar wind forces, since these provide
a pressure force similar to radiation pressure \citep{aug06}, and it
is known that stellar wind forces can be significantly stronger than
radiation forces for M stars \citep{pla05}.  While the smallest grain
size in the distribution may differ from our value of 0.9
\micron, this does not affect the ability of the model to fit the
observed emission spectrum with suitable modifications to the slope in
the size distribution and radius of the dust belt.

In Model 4 (dashed line of Fig.\ \ref{sed}), the dust grains are
assumed to lie at a range of distances from the star, but they are all
assumed to emit like black bodies, and so have temperatures $T=278.3\
L_*^{0.25}/\sqrt{r}$. The spatial distribution of the grains was taken
from the model of \cite{wya05b} in which dust is created in a
planetesimal belt at $r_0$, and then migrates inward due to
Poynting-Robertson (P-R) drag, but with some fraction of the grains
removed by mutual collisions on the way. In the model the dust ends up
with a spatial distribution which can be described by the parameter
$\eta_0$, such that the surface density falls off $\propto 1/(1+4 \
\eta_0 \ (1-\sqrt{r/r_0}))$. The emission spectrum could be fitted
using the parameters $r_0=30$ AU and $\eta_0=10$, resulting in dust
with temperatures upwards of 38 K. However, the density of the
planetesimal belt required to scale the resulting emission spectrum
showed that removal by collisions dominates over the P-R drag force in
such a way that $\eta_0$ should be closer to 1400 in this disk. Thus,
if this model is to have a true physical motivation, then we need to
invoke a drag force which is $\sim 140$ times stronger than P-R
drag. Such a force could come from the stellar wind, which causes a
drag force similar to P-R drag and which can be incorporated into the
model by reducing $\eta_0$ by a factor $1 + (dM_{\rm wind}/dt) \ c^2 /
L_*$ \citep{jur04}. Thus to achieve $\eta_0=10$ in this way we would
require the stellar wind to be $\sim 140$ times stronger than that of
the Sun. The high X-ray luminosity of TWA 7 \citep[$L_X = 9.2\pm1.0
\times 10^{29}$ erg/s,][]{ste00} may be indicative of a strong stellar
wind, since measured mass loss rates have been found to increase with
X-ray flux \citep{woo05}.  However the correlation found by
\cite[][see their Fig.\ 3]{woo05} breaks down at X-ray fluxes an order
of magnitude lower than that of TWA 7 (for which $F_X \sim 8 \times
10^6$ erg cm$^{-2}$ s$^{-1}$, and so it is not possible to use this
flux to estimate the mass loss rate with any certainty, and we simply
note that the mass loss rate required to achieve $\eta_0=10$ is not
incompatible with observations of mass loss rates of other stars. A
stellar wind drag force has also been invoked to explain structure in
the AU Mic disk \citep{str06,aug06}.

We note that we are not claiming that the radius, $r$, size
distribution, $q$, or parameter $\eta_0$ have been well constrained by
these fits. The SED does indicate that the disk around TWA 7 contains
grains at a range of temperatures. The fits of Figure \ref{sed}
illustrate two ways in which multiple temperatures in the disk may
arise from physically motivated models: the dust could have a range of
sizes, or it could be distributed over a range of distances. Other
models may also fit the data, including those in which dust has a
range of distances and sizes, and those in which the dust originates
in not one but multiple dust zones, as has been inferred for AU Mic
(Fitzgerald et al., in prep.).

The derived fits of Figure \ref{sed} do show differences in the mid-IR
spectrum. While this suggests that knowledge of the mid-IR spectrum
would enable us to distinguish between the two models, it is worth
pointing out that the exact shape of this spectrum is very sensitive
to the size distributions of small grains (in Model 3) and to the
radial distribution of grains interior to the planetesimal belt (in
Model 4). Thus it is possible that different assumptions about these
distributions could be made to provide a fit to the same observed
spectrum with both models. The only way to definitively break the
degeneracy is through imaging of the thermal emission from the
mid-IR to the submillimeter.  The single radius disk must look the
same at all wavelengths, while a radially distributed disk would look
larger at longer wavelengths, as more of the cold dust further from
the star is detected.

\subsection{Mass of the Disk}

\cite{low05} estimate the minimum mass in dust of the TWA 7 disk to be
$0.0033 \ M_{\rm lunar}$, under the assumption that the dust grain
size is 2.8 \micron.  In determining the mass of the disk from
submillimeter measurements, a key parameter is the temperature of the
dust.  Where possible, we discuss the derived mass for each of the
models discussed above.  For submillimeter fluxes, we can estimate the
mass of the disk directly for an assumed temperature and opacity from
the relation:

\begin{equation}
M_{\rm{disk}} = \frac{F_\nu \ d^2}{\kappa_\nu \ B_\nu(T_d)}
\label{mass}
\end{equation}

\noindent where $B_\nu(T_d)$ is the Planck function at the dust
temperature, $T_d$, and $\kappa_\nu$ is the absorption coefficient of
the dust. The derived mass is a strongly dependent function of the
value of $\kappa_\nu$.  For debris disk studies, a value of 1.7 cm$^2$
g$^{-1}$ is appropriate at 850 \micron\ \citep{den00}.  This is at the
upper end of the values derived by \cite{pol94}.

We can estimate the mass in Model 1 by using the submillimeter flux
predicted from the Rayleigh-Jeans tail of the model (2.4 mJy) as well
as the 80 K temperature and standard opacity. This gives a mass of
0.025 $M_\oplus$ ($\sim 2 \ M_{\rm lunar}$). This can be interpreted
as the mass of hot dust, with the proviso that the submillimeter
observation shows that there is more mass in colder dust as well.

Model 2 fits the submillimeter and 70 \micron\ emission well.  Under
the estimate of 45 K for the dust temperature of the disk, the
standard dust opacity and the measured 850 \micron\ flux, the TWA 7 disk
contains 0.2 $M_\oplus$ of material (18 $M_{\rm lunar}$).
Based on the uncertainty in the flux (random and systematic), the
uncertainty on the mass estimates are $\sim 30$\%.  The mass of the
TWA 7 disk is an order of magnitude greater than the mass of 0.011
$M_\oplus$ for the disk detected around AU Mic \citep{liu04} that is
also derived based on a single temperature fit to far-IR and
submillimeter data with $\kappa_\nu = 1.7$ cm$^2$ g$^{-1}$.

Model 3 contains dust grains at different temperatures based on their
size.  In this case, the size distribution of dust is well defined
(i.e., with known scaling), the total mass of dust is given by
$M_{{\rm tot}}=11 \ D_{{\rm max}}^{0.66}$, where $M_{{\rm tot}}$ is
the mass in Earth masses and $D_{{\rm max}}$ is in meters. While it is
impossible to observationally know $D_{{\rm max}}$ (since large
planetesimals are effectively invisible), in our model 95\% of the 850
\micron\ flux comes from grains $<0.4$ m.  This implies a mass of 6
$M_\oplus$, significantly higher than that of Model 2. In the TWA 7
system, $D_{{\rm max}}$ could be even larger (or smaller) than this,
so this discrepency carries little definitive weight.  The size
distribution is relatively shallow (i.e., there are lots of large
grains) which explains why the mass is much larger than in Model 2.

To derive a mass for Model 4 we assume an opacity of 1.7 cm$^2$
g$^{-1}$ and determine the mass of dust in the model required to
reproduce the observed 850 \micron\ flux, given that the dust has a
range of temperatures. The derived mass is almost exactly the same as
that at of Model 2 at $0.2 M_\oplus$. This is to be expected because
the submillimeter flux in both models is dominated by the coldest dust
at 38-45 K, and very little additional mass is required in Model 4 to
explain the mid-IR emission (as illustrated by the low mass in Model
1).

Given the inherent assumptions and unknowns in each of these models,
we adopt the results of the highly simple Model 2 as our most robust
estimate of the mass of the dust disk in TWA 7.  It depends on the
temperature of the grains producing the observed 850 \micron\ flux,
which are well fit by the cold dust model of Figure \ref{sed_1temp}.
The mass is also highly dependent on the value of $\kappa_\nu$, for
which we have adopted a value in line with other disk modeling work
\citep{den00}, and so can be compared with the disk masses derived by
other authors \citep[e.g.,][]{wya03,liu04,naj05}.

\section{Discussion}
\label{disc}

\subsection{Disks in the TW Hydrae Association}

Recent MIPS data from the Spitzer Space Telescope show that most stars
in the TWA show no evidence of circumstellar dust out to 70 \micron\
\citep{low05}.  These results confirm the assertion of \cite{wei04}
and \cite{gre03} that there is a bimodal distribution in the TWA:
either stars have strong excesses associated with warm dust emission,
or they have very weak emission consistent with very cold disks.  The
exceptions are the proto-planetary disks around TW Hya and Hen 3-600,
which both show evidence of warm and cold disk components
\citep{wil00,zuc01}.

\begin{deluxetable*}{lccccrrcl}[ht!]
\tablecolumns{9} 
\tablewidth{0pc} 
\tablecaption{Fluxes and Masses of Disks in the TWA}
\tablehead{\colhead{Star} & \colhead{Sp.\ Type} & \colhead{Gas?} &
  \colhead{Optical Depth} & \colhead{Flux} &
\colhead{$\lambda$} &  \colhead{Mass} & \colhead{Temp} & \colhead{Reference}  \\
& & & & \colhead{[mJy]} & & \colhead{[$M_{\rm{lunar}}$]} & \colhead{[K]} & }
\startdata 
HR 4796A & A0 & no & thin & $19.1 \pm 3.4$ & 850 \micron & 19 & 99 & \cite{she04}  \\ 
HD 98800 & K5 & no & thick & $111.1 \pm 0.01$ & 800 \micron$^a$ & 28 & 150 &
\cite{pra01} \\
TW Hya & K7 & yes & thick & $8 \pm 1$ & 7 mm & $8 \times 10^5$ &  -- & \cite{wil00}, SED fit \\
TWA 7 & M1 & no & thin & $9.7 \pm 1.6$ & 850 \micron & 18 & 45 & this work \\
TWA 13 & M1 & no & thin & $27.6 \pm 5.9$ & 70 \micron & $> 0.0019$ &  65 & \cite{low05} \\
Hen 3-600 & M3 & yes & thick & $\sim 65$ & 850 \micron & 304$^b$ & 20$^c$ & \cite{zuc01} \\ 
\enddata
\tablecomments{$^a$ $\kappa_{\rm 850 \micron}$ adjusted by
  $1/\lambda$. $^b$ mass estimate based on conservative value of
  $\kappa_{\rm 850 \micron} = 1.7$ \opacity; $^c$ temperature
  constraint on cold dust component only. \vspace*{0.3cm}}
\label{massTWA}
\end{deluxetable*}

Lower limits on warm dust emission were set for all TWA members
(except TW Hya) by \cite{wei04}.  The absence of dust near the star is
often taken as a signature of a centrally depleted debris, rather than
an accreting, proto-planetary disk.  There are a few main sequence
stars for which warm dust is present (i.e., HD 69830 and $\eta$
Corvi), but this may be a transient phenomenon \citep{wya07}.  Based
on their data, \cite{wei04} deduced that the non-detections implied an
absence of material in terrestrial planet region and that, except for
TW Hya, there were no long-lived disks in the association which could
still form planets.  An outstanding question is whether the stars
without warm disks have potential disk material locked up in
undetected planets, which would imply that dusty systems represent
failed planetary systems, or whether non-detections represent disk
systems in which the dust is too cold to be detected in the
mid-IR. Only large-scale searches for cold dust around a statistically
significant number of stars can clarify whether a sizable population
of disks exist which are too cold to detect at mid-IR wavelengths.
Such a survey is planned using the new SCUBA-2 camera at the JCMT
\citep{mat07}.

The disks now known in the TWA each have very different
properties. The disk around TW Hya (K7) still contains molecular gas
\citep{kas97}, meaning it is proto-planetary or in a transition from a
proto-planetary to a debris disk. It is a broad, face-on disk which
extends to 135 AU with evidence of a dip in flux at 85 AU
\citep{kri00,wil00,tri01,wei02,qi04}.  Hen 3-600 also shows evidence
of hosting an accreting, gas-rich disk \citep{muz00}.  The HR 4796A
(TWA 11) disk contains a narrow dust ring at 70 AU from the star
\citep{jay98,koe98,sch99,tel00} and no detectable gas in emission line
studies \citep{gre00} or more recent searches for absorption due to
circumstellar gas along the line of sight \citep{che04}, a technique
which is highly dependent on the temperature profile of the gas
because the disk is not edge-on.  In both the Hen 3-600 and HD 98800
systems, the disk orbits only one member of the binary
\citep{jay99,geh99} with evidence for cooler dust in circumbinary
orbits \citep{zuc01}.  In fact, HD 98800 is a quadruple system, so the
dust is in a circumbinary orbit around a spectroscopic binary.

The discussion of the TWA disk population can be illuminated by the
results of recent Spitzer study of the Upper Sco Association (with an
estimated age of $3-5$ Myr) by \cite{car06}.  Their MIPS observation
found optically thin disks around A-type stars, no disks around
solar-like stars, and optically thick, accreting disks around K- and
M-type dwarfs, suggesting that disk evolution proceeds more quickly
around higher mass stars.  A similar result was found for the H and
$\chi$ Persei double cluster at 13 Myr \citep{cur07}.  The sample of
stars in Upper Sco was 204 stars with 31 detections at 8 \micron, well
distributed across spectral types. In the TWA, we have disk detections
around only a handful of members, and the association is much smaller,
with only 18 members.  However, both the optically-thick, gas-rich
disks in the TWA are hosted by K and M stars, whereas the only A star
with a disk hosts an optically-thin debris disk.  We note that HD
98800 is noted to be gas-poor, but with an optically-thick dust disk
\citep{wei00,zuc93}.

The masses and temperatures of the disks are compared for the TWA
members with disks in Table \ref{massTWA}.  All six disks are detected
and their SEDs modeled in the recent paper by \cite{low05}.  Where
possible, the masses in Table \ref{massTWA} are derived from
submillimeter fluxes or fits to SEDs, rather than infrared values
which provide lower limits only.  This is only an issue for TWA 13
which has not yet been detected at submillimeter or millimeter
wavelengths.  Scaling $\kappa_{\rm 850 \micron} = 1.7$ \opacity\
to 70 \micron\ implies a lower mass limit of $0.1 \ M_{\rm lunar}$
based on cold grains.

Of the debris disks, the most massive is the disk around the HD 98800
system, but its mass is comparable to that of the disk around the
earlier star HR 4796A.  Based on submillimeter masses, the TWA 7 disk
is roughly comparable to that of HR 4796A.  It is clear that we cannot
yet identify a systematic trend in mass or evolutionary phase with
spectral type in the TWA.  We note that, given the presence of massive
disks around components of the multiple systems HD 98800 and HR 4796,
their dust disk lifetimes do not appear to be any shorter than that
around a single star.

\subsection{Disks Around Late-type Stars}

Table \ref{masslate} shows the compilation of the few known debris
disks around late-type (K and M) stars.  Although a debris disk has
been historically claimed around HD 233517 \citep{syl89}, \cite{jur06}
conclude that it is a giant, not a main sequence, star, and so we do
not include it.  Similarly, we exclude HD 23362, although it has a
measured excess, because \cite{kal02} attribute the emission to a
uniform surrounding dust cloud, not a debris disk, around a K2III star
at 187 pc distance.  All fluxes are measured at 850 \micron\ with
SCUBA, except where noted. For one of these stars, only an upper limit
is observed in the 850 \micron\ flux.  Of the eight solid detections,
two (HD 92945, TWA 13) are at a single wavelength in the mid-infrared,
and two (GJ 182 and GJ 842.2) are detected only in the submillimeter.
For the four remaining disks (HD 53143, $\epsilon$ Eri, AU Mic and TWA
7), there is a trend of increasing mass with {\it later} spectral
type, but it must be noted that the mass dependence could also be
attributed to the known trend of declining dust masses around older
stars \citep{rhe06} in the cases of HD 53143 and $\epsilon$ Eri since
they are significantly older than AU Mic and TWA 7.  The youngest
star, TWA 7, has the most massive disk.  We are obviously also biased
toward more massive disks at larger distances.  AU Mic's disk (9.9 pc)
could not have been detected in the observation which detected TWA 7's
disk at 55 pc.

\begin{deluxetable*}{lcccrrrcl}
\tablecolumns{9} 
\tablewidth{0pc} 
\tablecaption{Masses and Mass Limits of Debris Disks around Late-type
  Main Sequence Stars}
\tablehead{\colhead{Star} & \colhead{Spectral} &
  \colhead{Distance} & \colhead{Age} & \colhead{Flux} &
  \colhead{$\lambda$} & 
  \colhead{Mass} & \colhead{Temp} & \colhead{Mass Reference}  \\
& \colhead{Type}& \colhead{[pc]} & \colhead{[Myr]} & \colhead{[mJy]} &
   & \colhead{[$M_{\rm{lunar}}$]} & \colhead{[K]} & }
\startdata 
HD 69830$^a$ & K0 & 12.6 & 600-2000 & $< 7$ & 850 \micron & $<0.24$ & 100 & this
work\\
HD 92945 & K1 & 22 & 20-150 & 271 & 70 \micron & $> 0.002$ & 40 & \cite{che05} \\
HD 53143 & K1 & 18 & 1000 & $82.0 \pm 1.1$ & 30-34 \micron & $> 6.5
\times 10^{-6}$ & $120 \pm 60$ & \cite{che06} \\
& & & & -- & optical & $> 0.0096$ & 60$^b$ & \cite{kal06} \\
$\epsilon$ Eri & K2 & 3 & $730 \pm 200$ & $40 \pm 1.5$ & 850 \micron &
0.1  & 85 & \cite{she04} \\
GJ 842.2 & M0.5 & 20.9 & 200 & $25 \pm 4.6$ & 850 \micron & $28 \pm 5$ & 13 &
\cite{les06} \\ 
GJ 182$^c$ & M0.5 & 26.7 & $100^{+50}_{-30}$ & $4.8 \pm 1.2$ & 850
\micron & $ > 2.1$ & 40 + 150$^d$ & \cite{liu04} \\ 
AU Mic$^c$ & M1 & 9.9 & 10 & $14.4 \pm 1.8$ & 850 \micron & 0.89 & 40 &
\cite{liu04} \\ 
TWA 7$^c$ & M1 & 55 & 8 & $9.7 \pm 1.6$ & 850 \micron & 18 & 45 & this work \\ 
TWA 13$^c$ & M1 & 55 & 8 & $27.6 \pm 5.9$ & 70 \micron & $> 0.0019$ & 65 & \cite{low05} \\ 
\enddata 
\tablecomments{$^a$ We have used the adopted temperature of
100 K but revised downward the estimated flux and mass based on
re-analysis of data originally presented in \cite{she04}. \cite{wya07}
suggest that the warm dust around this source must be transient. $^b$
Temperature from \cite{zuc04}. $^c$ Ages derived from association
membership. $^d$ two-component fit. \vspace*{0.5cm}}
\label{masslate}
\end{deluxetable*}

Spitzer has made great progress in the last year detecting infrared
excess from main sequence stars \citep[see the review by][]{wer06}.
However, not many of these have been around late-type stars, and for
those detections which have been made
\citep{gor04,bei05,uzp05,bry06,smi06}, no estimates of mass in the
disk exist.  The detections are typically toward field stars or very
distant targets in the galactic plane, although one (P922) is a member
of the cluster M47 \citep{gor04}.  We do not list these candidates in
Table \ref{masslate}.  One exception is the excess around HD 92945
(K1V), for which \cite{che05} measure a minimum disk mass of $2\times
10^{-3} \ M_{\rm lunar}$.

As discussed in $\S$ \ref{3.1}, the degeneracies in the models are
best broken with thermal imaging.  Of the disks compiled in Table
\ref{masslate}, only $\epsilon$ Eri (3.3 pc) has been well resolved at
submillimeter wavelengths.  The distance of the TWA makes imaging of
100 AU ($\sim$ 2\arcsec) scale disks impossible with single dish
telescopes in the submillimeter.  $\epsilon$ Eri would be exceedingly
difficult to map if it were even three times more distant.  We will
have to rely on arrays with higher sensitivity to obtain maps like
that of $\epsilon$ Eri around most low mass stars unless many more are
discovered within 10 pc.  In the long-term, mid-IR images will be
possible with the planned MIRI instrument on the James Webb Space
Telescope.  In the short-term, submillimeter imaging will be possible
with the Atacama Large Millimeter-submillimeter Array (ALMA). Far-IR
observations will be possible with the Herschel Space Observatory,
although imaging of 100 AU disks will only be possible for stars
within 10 pc.

\section{Summary}
\label{conc}

We have detected submillimeter excess emission arising
from the dust disk around TWA 7 at 450 and 850 \micron\ using SCUBA on the
JCMT.  Based on our photometry and recent data from Spitzer, we derive
a disk mass of 0.2 $M_\oplus$ (18 $M_{{\rm lunar}}$) for a temperature
of 45 K.  This model effectively fits the 70, 450 and 850 \micron\
data with a blackbody.  To fit these data and the 24 \micron\ flux
requires dust at a range of temperatures, and we show that this could
arise from dust at one radius with a range of sizes, or from dust of
one size at a range of distances from the star.  Based on the SED
alone, it is not possible to determine which physical model is
dominant.

While the multiple system HD 98800 appears to harbour the most massive
debris disk in the TWA, disks of relatively comparable masses are
observed around the A0 star HR 4796A and the M1 star TWA 7.  Therefore, the
formation of debris disks does not appear to 
be solely a function of the mass of the parent star.  A comparison of
masses of disks in the TWA reveals no trend in mass or evolutionary
state (gas-rich, proto-planetary vs.\ debris) as a function of
spectral type, although the detection of proto-planetary disks around
the latest stars is consistent with the results of \cite{car06} toward
the Upper Sco Association and \cite{cur07} in the double cluster H and
$\chi$ Persei.

\cite{kal06} came to the same conclusion with regard to other
debris disks. They surmise that nurture could explain the presence or
absence of disks at later epochs.  If the environment dynamically
heats the disk such that the large planets fail to form, then dust
remains for a longer timescale.  The dynamically quiet systems then
may quickly form planets, leaving no disk to be observed at later
epochs, although there is as yet no evidence for any
correlation between stars with debris and/or planets \citep{mor06}.

\acknowledgements

The authors acknowledge our anonymous referee for an insightful and
constructive report.  As well, the authors thank B.\ Zuckerman for
providing us with the previous 850 \micron\ flux measurement from the
thesis of R. Webb, and P.\ Smith for providing the stellar fit to TWA
7 of \cite{low05} to maximize consistency with their analysis. We also
acknowledge useful conversations with P.\ Hauschildt and R.\ Gray
regarding the spectra of M dwarfs. We thank our telescope operator E.\
Lundin and the staff at the JCMT for their support.  B.C.M
acknowledges support of the National Research Council of Canada
through a Plaskett Fellowship.  P.K. acknowledges support from
GO-10228 provided by STScI under NASA contract
NAS5-26555. M.C.W. acknowledges support of the Royal Society.

\end{document}